\documentclass[aps,prb,reprint,superscriptaddress]{revtex4-1}

\usepackage{amsmath,amssymb}
\usepackage{graphicx}
\usepackage{dcolumn}
\usepackage{bm}

\usepackage[utf8]{inputenc}
\usepackage[T1]{fontenc}
\usepackage{mathptmx}
\usepackage{epstopdf,xcolor}

\begin{document}


\title{Origin of Irradiation Synergistic Effects in Silicon Bipolar Transistors: a Review}



\author{Yu Song}
\email{kwungyusung@gmail.com} 
\affiliation{College of Physics and Electronic Information Engineering, Neijiang Normal University, Neijiang 641112, China}

\author{Su-Huai Wei}
\email{suhuaiwei@csrc.ac.cn}
\affiliation{Beijing Computational Science Research Center, Beijing 100193, China}


\date{\today}

\begin{abstract}
The practical damage of silicon bipolar devices subjected to mixed ionization and displacement irradiations is usually evaluated by the sum of separated ionization and displacement damages. However, recent experiments show clear difference between the practical and summed damages, indicating significant irradiation synergistic effects (ISEs). Understanding the behaviors and mechanisms of ISEs is essential to predict the practical damages.
In this work, we first make a brief review on the state of the art,
critically emphasizing on the difficulty encountered in previous models to understand the dose rate dependence of the ISEs. We then introduce in detail our models explaining this basic phenomenon, which can be described as follows. Firstly, we show our experimental works on PNP and NPN transistors. A variable neutron fluence and $\gamma$-ray dose setup is adopted. Fluence-dependent `tick’-like and sublinear dose profiles are observed for PNP and NPN transistors, respectively. Secondly, we describe our theoretical investigations on the positive ISE in NPN transistors. We propose an atomistic model of transformation and annihilation of displacement-irradiation-induced $\rm V_2$ defects in p-type silicon under ionization irradiation, which is totally different from the traditional picture of Coulomb interaction of oxide trapped charges in silica on charge carriers in irradiated silicon. The predicted novel dose and fluence dependences are fully verified by the experimental data. Thirdly, the mechanism of the observed negative ISE in PNP transistors is investigated in a similar way as in the NPN transistor case.
The difference is that in n-type silicon, displacement-irradiation-induced VO defects also undergo an ionization-induced transformation and annihilation process.
Our results show that, the evolution of displacement defects due to carrier-enhanced defect diffusion and reaction is the dominating mechanism of the ISEs. Finally, we give a perspective on future investigations on the ISEs when the displacement and ionization irradiations are present simultaneously.
\end{abstract}



\maketitle

\section{Introduction}

{\bf Irradiation {synergistic} effect.}
Silicon (Si)-based NPN and PNP transistors are building blocks of modern {microelectronics}, such as operational amplifiers and voltage comparators.
When these devices are used in outer space and other extreme environments such as nuclear reactors, both charge carriers and atomic displacements will be generated in the silica and silicon materials, which result in ionization and displacement damages in the devices.~\cite{oldham2003total,srour2003review}
Conventionally, the practical damage is usually assumed as a simple sum of the ionization damage (ID) and displacement damage (DD), which then can be evaluated individually at ground by carrying out $\gamma$-ray and neutron irradiation experiments, respectively.
The primary reason for the separability of the damages is that, the ID is caused by the ionization defects, oxide trapped charges (OTs) and interface traps (ITs), generated in the silica and near interface,~\cite{Dressendorfer1998basic} while the DD is caused by the displacement defects, such as divacancies (V$_2$) and vacancy-oxygen complex (VO), generated in Si.~\cite{fleming2010annealing}
Accordingly, it was widely and always believed that there is no direct interaction between these two irradiation-induced defect dynamics and the practical damage can be separated.

However, in recent experiments it has been clearly demonstrated that, when these bipolar transistors are simultaneously or subsequently irradiated by displacement and ionization irradiations, the practical damage is usually smaller or larger than the simply summed damage.~\cite{Barnaby2001proton,Barnaby2002Analytical,gorelick2004effects,
Li2012Synergistic,Li2012Simultaneous,Li2015SynergisticEffect,Wang2016Ionizing,
song2019mechanism,song2020defect}
{These phenomena} can be called as irradiation synergistic effects (ISEs), which are negative in PNP transistors and positive in NPN transistors, respectively.
The presence of ISEs imply that one cannot evaluate the practical damage by {carrying} out separated neutron and $\gamma$-ray irradiations and simply summing the damages up. To obtain a relatively true damage, one must know the behaviors of the ISEs and understand the underlying mechanisms.

\begin{figure}[!b]
\centering
\includegraphics[width=\linewidth]{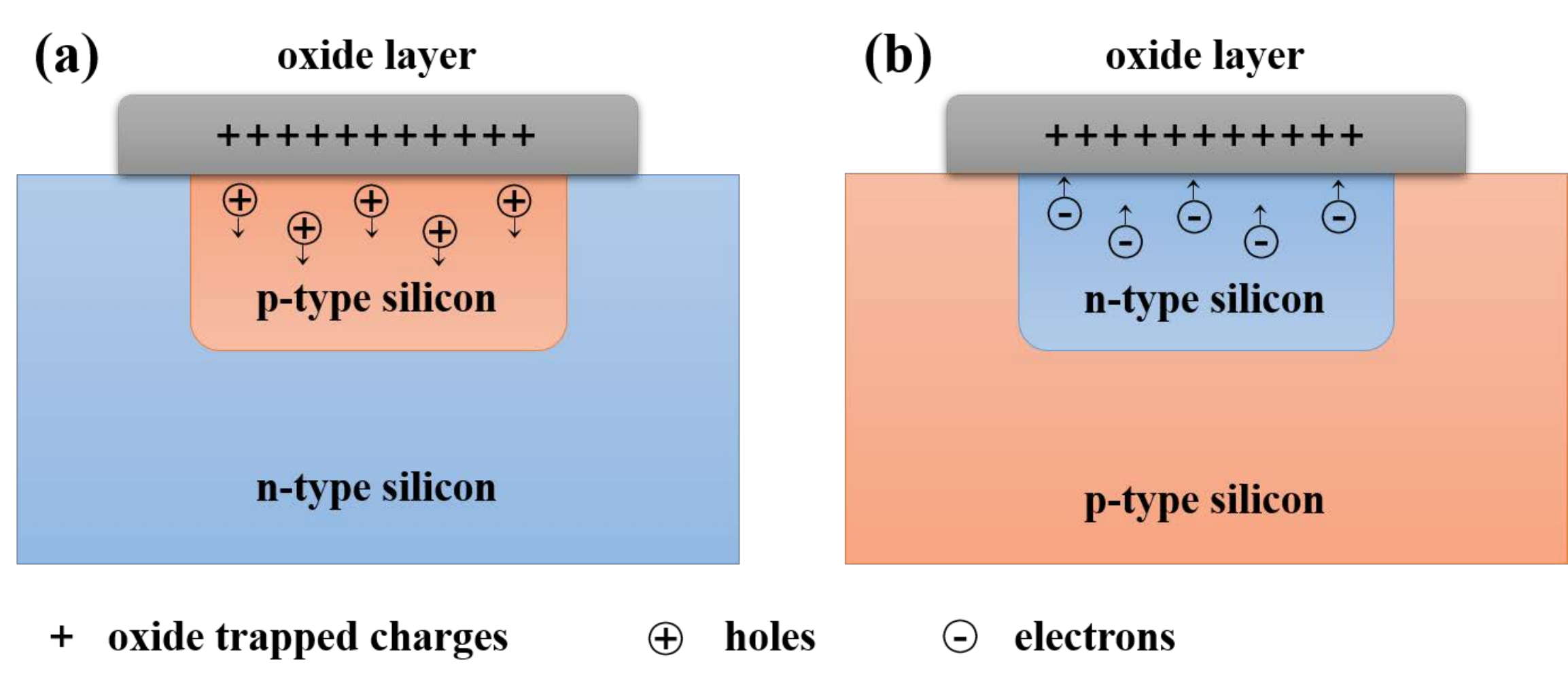}
\caption{Schematic diagram of the Coulomb interaction mechanisms for NPN (a) and PNP (b) transistors. The short arrows indicate the drift of the charge carriers due to the influence of OTs in the oxide layer.
}\label{fig:coulomb}
\end{figure}

{\bf {Existing model of the ISEs}.}
Currently, the ISEs are explained in a point of view of interfacial interaction in the device scale, see, Fig. 1. Especially, the effect is believed to stem from the Coulomb interaction of OTs in the silica layer on the charge distributions in the displacement-damaged silicon base region, where the non-radiative Shockley-Read-Hall (SRH) carrier recombination occurs.~\cite{Barnaby2001proton,Barnaby2002Analytical}
To be fair, this point of view is rather natural: the outcomes of the ID and DD are the ionization and displacement defects in the silica layer and silicon region, respectively, so, intuitively, the synergistic effect of the ID and DD should come from the Coulomb interaction between the charged defects and carriers in these two different regions. Actually, the opposite signs of the ISEs in NPN (which is $+$) and PNP (which is $-$) transistors can be consistently explained by this mechanism, as followings.
A positive ISE arises in NPN transistors because the positively charged OTs repel the positively charged holes as majority carriers near the surface of the p-type silicon base, which reduces the difference of hole and electron densities and enhances the SRH recombination current in the base region, see Fig. 1(a). 
On the contrary, a negative ISE would arise in PNP transistors because the positive OTs attract the negatively charged electrons as majority carriers near the surface of the n-type silicon base, which widens the difference of carrier densities and thus suppresses the SRH recombination in the base region, see Fig. 1 (b).

{\bf Difficulty of the existing model in understanding the ISE.} However, we notice that some of the experimental observations of the ISEs cannot be explained by the existing mechanisms. The most remarkable feature is the dose-rate dependence of the effect. According to a reduced high-dose-rate sensitivity,~\cite{pease2008eldrs} the OTs will become fewer for decreasing dose rate irradiation at a fixed total dose.~\cite{song2020universal} According to the Coulomb interaction mechanism, the ISE should become weaker as the dose rate {decreases}. However, this deduction is totally opposite to the experimental results, where the synergistic effects between DD and ID are found to be enhanced at lower dose rate $\gamma$-ray irradiations.~\cite{song2019mechanism,song2020defect}
On the other hand, the ionization dose (dose rate) and displacement fluence/flux dependencies of the ISEs are essential for the understanding and modeling. However, existing experimental works usually use proton and electron irradiations, where the simultaneously generated DD and ID are hard to separate.~\cite{Barnaby2001proton,Li2010Combined} Meanwhile, existing investigations usually consider fixed ionization dose or displacement fluence. So, till now, the behavior of the ISEs as functions of the ionization dose, dose rate, and displacement fluence is still not clear.

To sum up, the ISEs are important to evaluate practical damages of silicon bipolar transistors in irradiation environments. However, there are still huge gaps in the investigations of the behaviors and mechanisms of the effects. The behavior of the ISEs is far from clear due to the fixed displacement fluence and/or ionization dose in existing experiments. The understanding of the effect is based on a point of view of an interfacial interaction, which seems natural but cannot explain the dose rate dependence of the effect. Thus, new experimental studies and mechanisms that can self-consistently describe the experimental observations are urgently needed.

\begin{figure}[!t]
\centering
\includegraphics[width=0.86\linewidth]{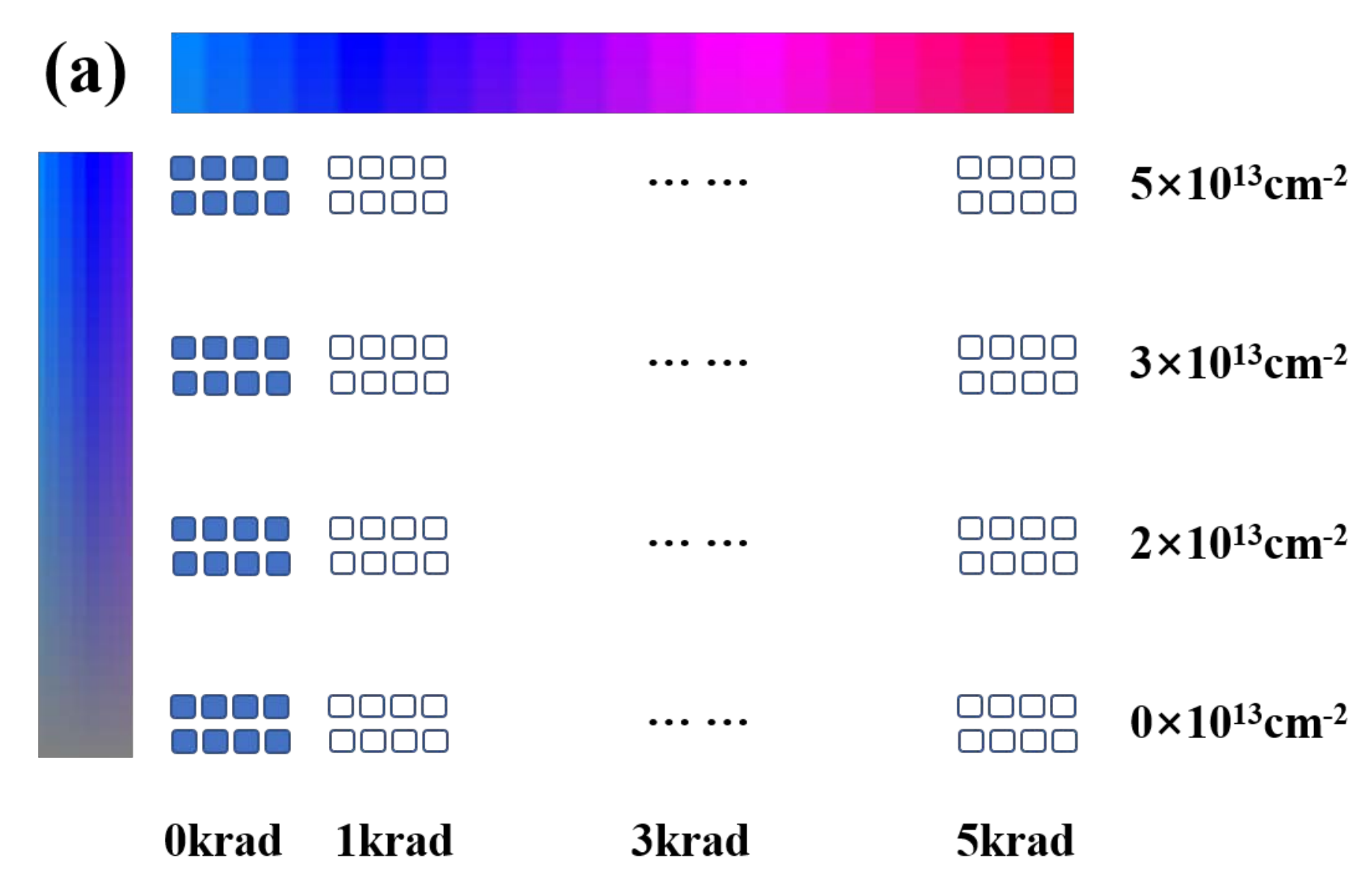}\\
\includegraphics[width=0.86\linewidth]{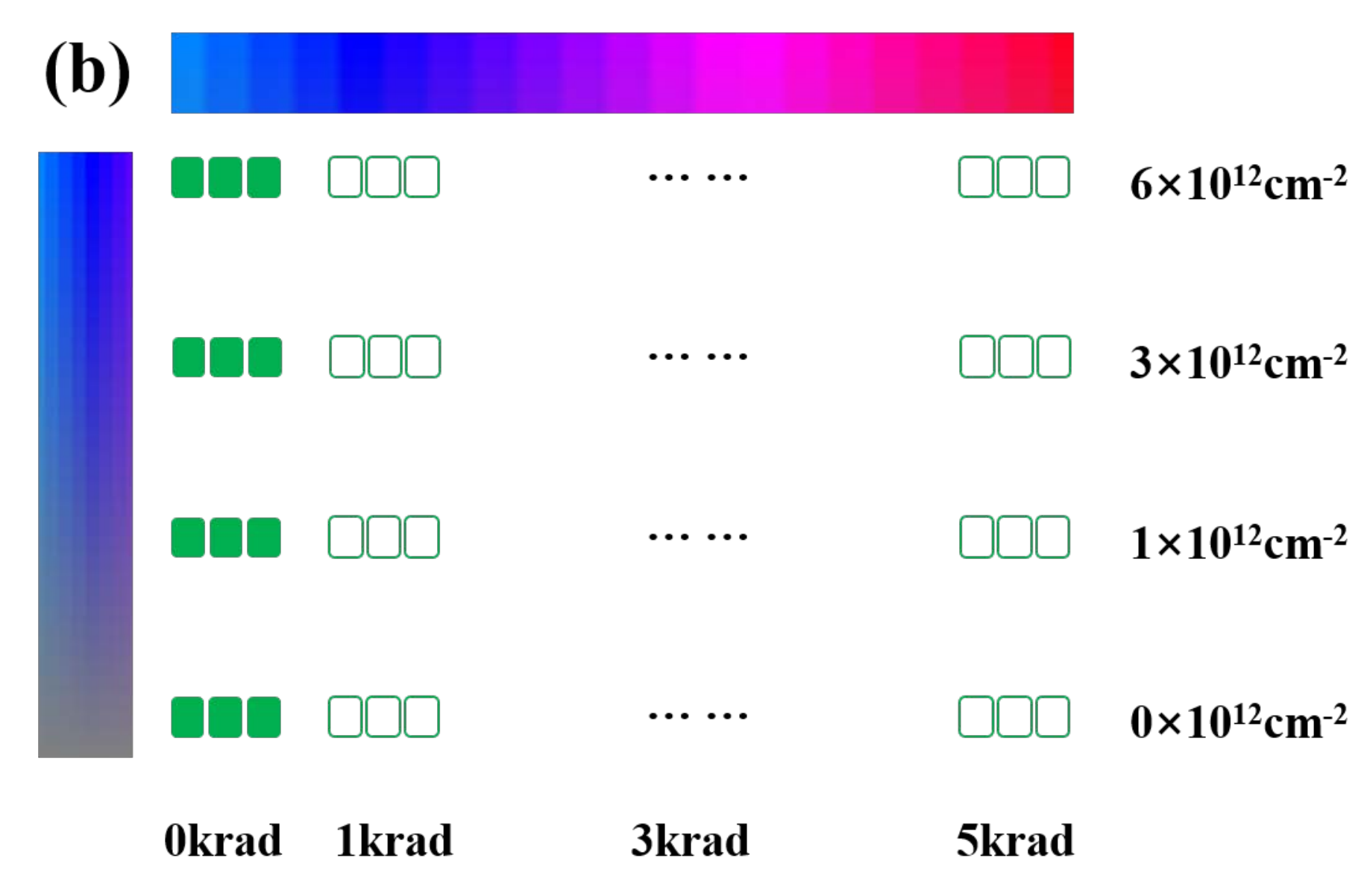}
\caption{Setup of the neutron/$\gamma$-ray irradiation experiments for PNP (a) and NPN (b) transistors. For each of the irradiation conditions, {8 PNP transistors or 3 NPN transistors} are used. The color stripes schematically indicate the strength of damages.
}\label{fig:expflow}
\end{figure}

\section{Behaviors of the ISEs}

\subsection{Experimental setup to explore the behaviors}

To explore the behavior of the ISEs in PNP and NPN transistors,~\cite{song2019mechanism,song2020defect} LM324N chips with 4 PNP-type input-stage transistors in each chip and 3DK9D NPN-type transistors were selected as the research object, respectively.
These devices were chosen because they are sensitive to both ID and DD.
The neutron irradiation hardly influences the dynamics of OTs in the oxide layer,~\cite{raymond1987comparison}
hence, we adopt a sequential neutron/$\gamma$-ray irradiation configuration.
The processes of the sequential irradiation experiments for PNP transistors is shown in Fig.~\ref{fig:expflow} (a). 
The transistors are first irradiated by neutron to the fluence as indicated in the figure along the vertical direction; then they are irradiated by $\gamma$-ray to the dose as indicated in the figure along the horizontal direction. 
By these irradiations the practical damages can be obtained.
The summed damages as a reference are obtained by another two pure $\gamma$-ray irradiations.
For each neutron-$\gamma$-ray condition, 8 PNP transistors (within 2 chips) are used to avoid possible misjudgment due to sample variability.
Note, existing experiments~\cite{Barnaby2001proton,
gorelick2004effects,
Li2012Synergistic,
Wang2016Ionizing}
usually adopt only a fixed displacement {fluence} and ionization dose, hence is very different from the present experimental setup.
The radiation sources, irradiation bias, and testing systems are described in detail in Ref. \onlinecite{song2019mechanism}.
The ISE of NPN transistors is investigated by a very similar sequential neurton/$\gamma$-ray irradiation experiment, which is shown in Fig.~\ref{fig:expflow} (b) and has been described in detail in Ref.~\onlinecite{song2020defect}.

\begin{figure}[!t]
\centering
\includegraphics[width=0.9\linewidth]{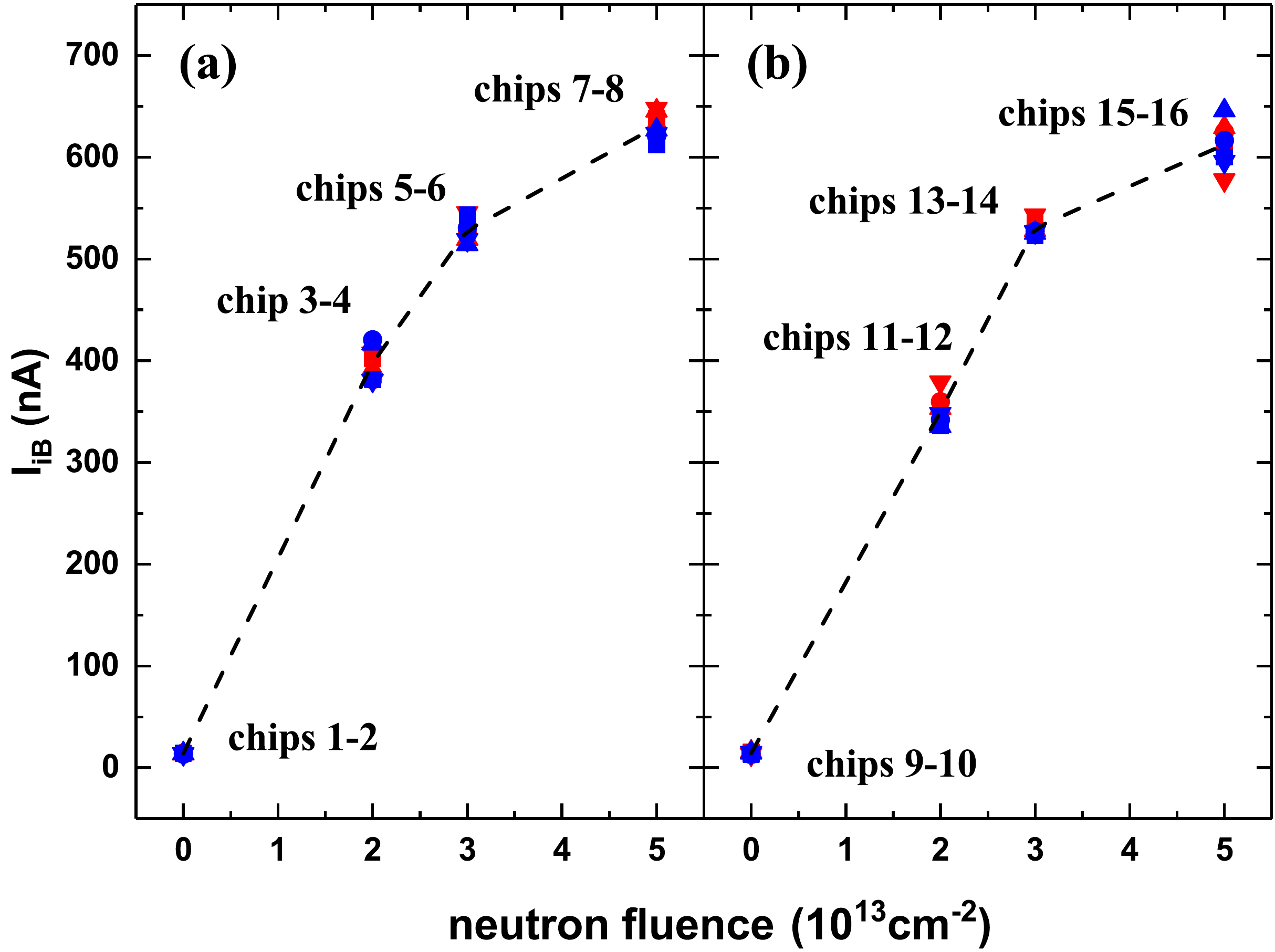}
\caption{The neutron fluence dependence of the input bias current of LM324N chips. 
For each fluence, the two chips are shown by red and blue, respectively; in each chip, the four transistors are shown by different symbol shapes.
Reproduced from Song, Y. \emph{et al.}, \emph{ACS Applied Electronic Materials} \textbf{2019}, \emph{1}, 538–547. Copyright 2019 American Chemical Society.
}\label{fig:DDradiation}
\end{figure}

\begin{figure*}[!t]
\centering
\includegraphics[width=\textwidth]{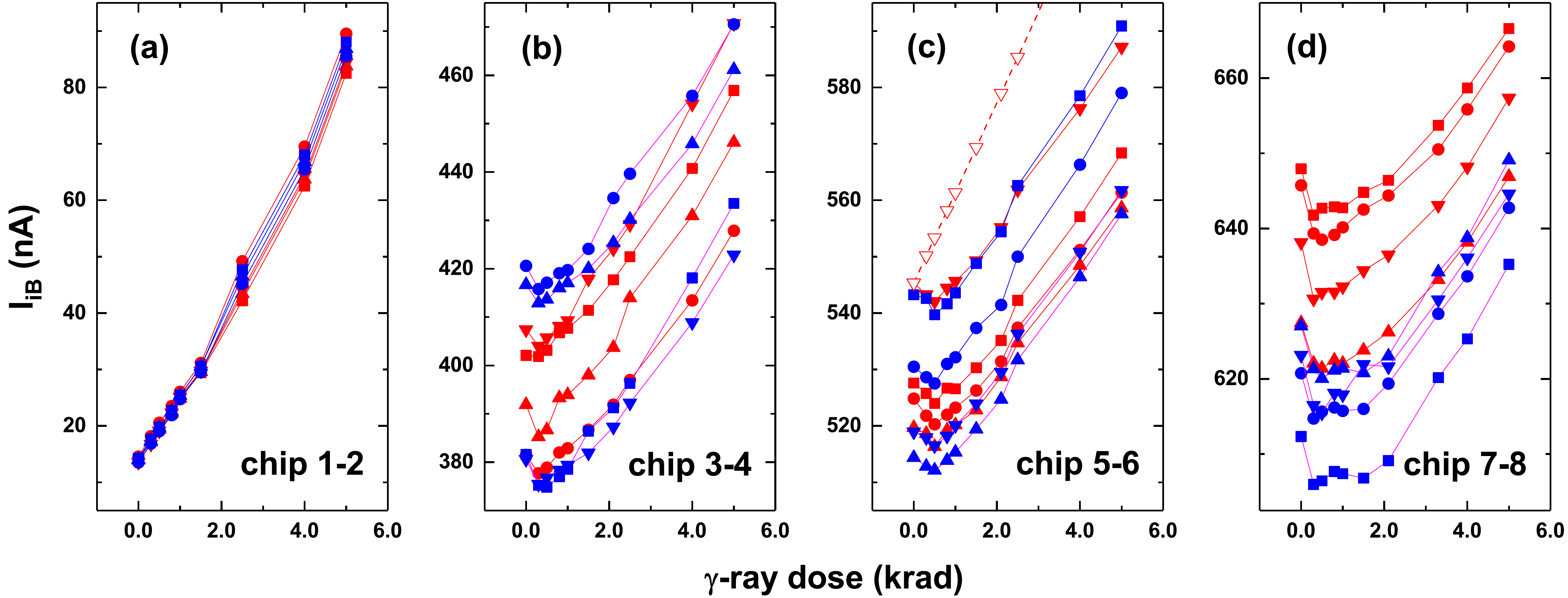}
\\
\caption{The $\gamma$-ray dose dependence of the input bias current of LM324N chips for initial neutron fluence of $0$ (a), $2\times 10^{13}~\text{cm}^{-2}$ (b), $3\times 10^{13}~\text{cm}^{-2}$ (c), and $5\times 10^{13}~\text{cm}^{-2}$ (d), respectively. The dose rate is $2.2~\text{mrad(Si)/s}$. In each split, the eight transistors are shown by different colors and symbol shapes as in Fig. 3.
Reproduced from Song, Y. \emph{et al.}, \emph{ACS Applied Electronic Materials} \textbf{2019}, \emph{1}, 538–547. Copyright 2019 American Chemical Society.
} \label{fig:TIDafterDDradiation}
\end{figure*}

\subsection{Fluence-dependent `tick'-like dose curves of PNP transistors}

{\bf Artificial summed damages.}
The individual DDs of the PNP transistors are shown in Fig. 3, from which a sub-linear dependence of the input bias current ($I_{iB}$) on the neutron fluence is clearly seen. 
It is also seen that, the DDs of the 8 transistors under a same fluence are different due to sample-to-sample variability. 
To obtain the artificial summed damages, Nos. 1-2 and 9-10 chips were irradiated by a low and high dose rate $\gamma$ rays, respectively. The results for the former are displayed in Fig. 4 (a).
An almost linear increase of the ID is seen and an average slope of $k_0^L = 15.6$ nA/krad(Si) and $k_0^H = 8.0$ nA/krad(Si) is extracted for the low and high dose rate case, respectively.
The summed damage reads:
\begin{equation}
\Delta I_{iB}^{D+I} = D_0 +k_0^{L(H)} x,
\end{equation}
where $D_0$ is the initial DD before $\gamma$-ray irradiations. A typical result is plotted in Fig. 4 (c) as the dashed line.

\begin{figure}[!b]
\centering
\includegraphics[width=0.95\linewidth]{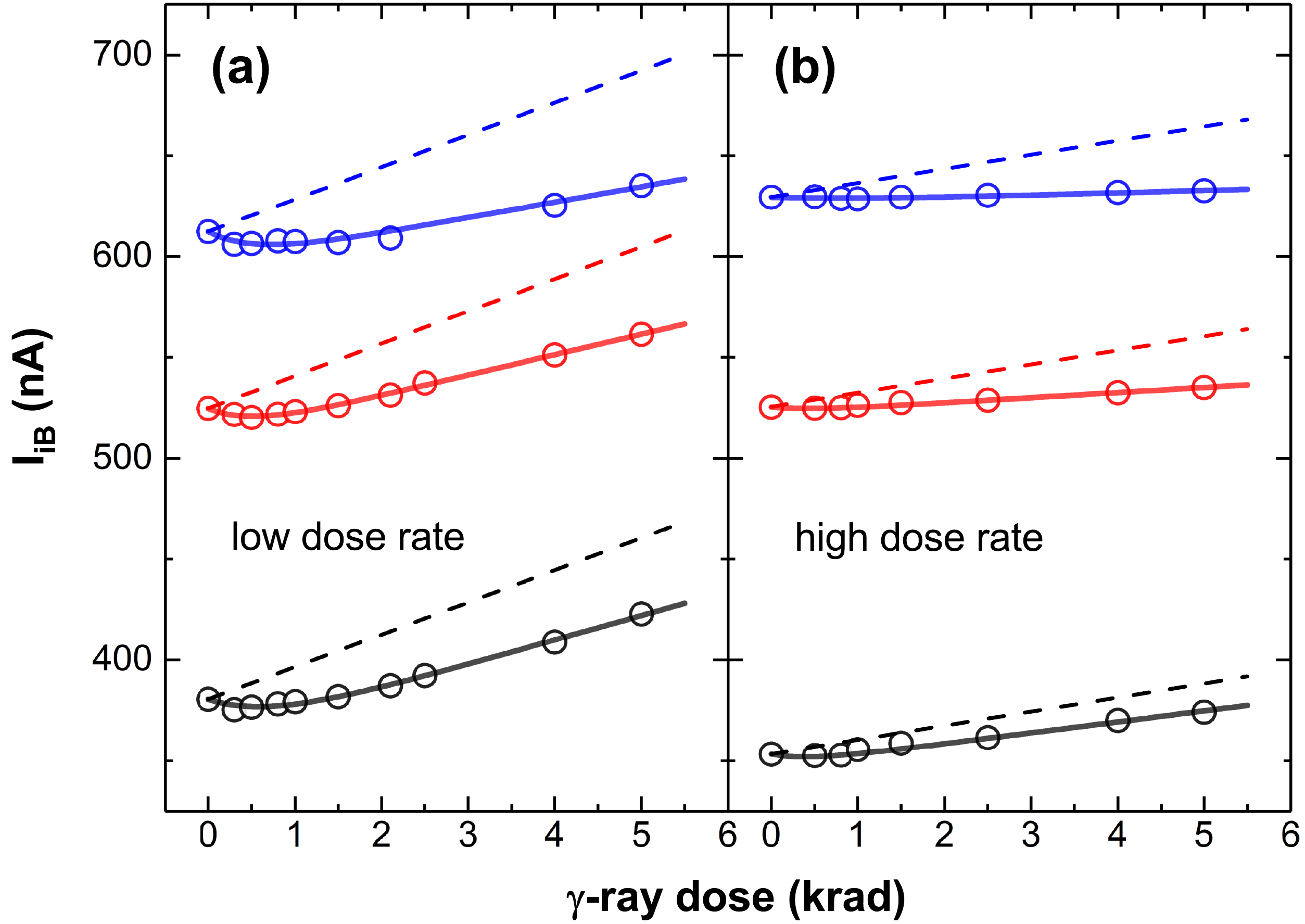} 
\caption{The typical experimental data (dots) and fitting curves (solid) of the dose dependence of the input bias current of PNP-type LM324N for low (a) and high (b) dose rate, respectively. The dashed lines show the summed damages. 
Reproduced from Song, Y. \emph{et al.}, \emph{ACS Applied Electronic Materials} \textbf{2019}, \emph{1}, 538–547. Copyright 2019 American Chemical Society.
} \label{fig:TIDpure}
\end{figure}

\begin{figure}[!t]
\centering
\includegraphics[width=0.85\linewidth]{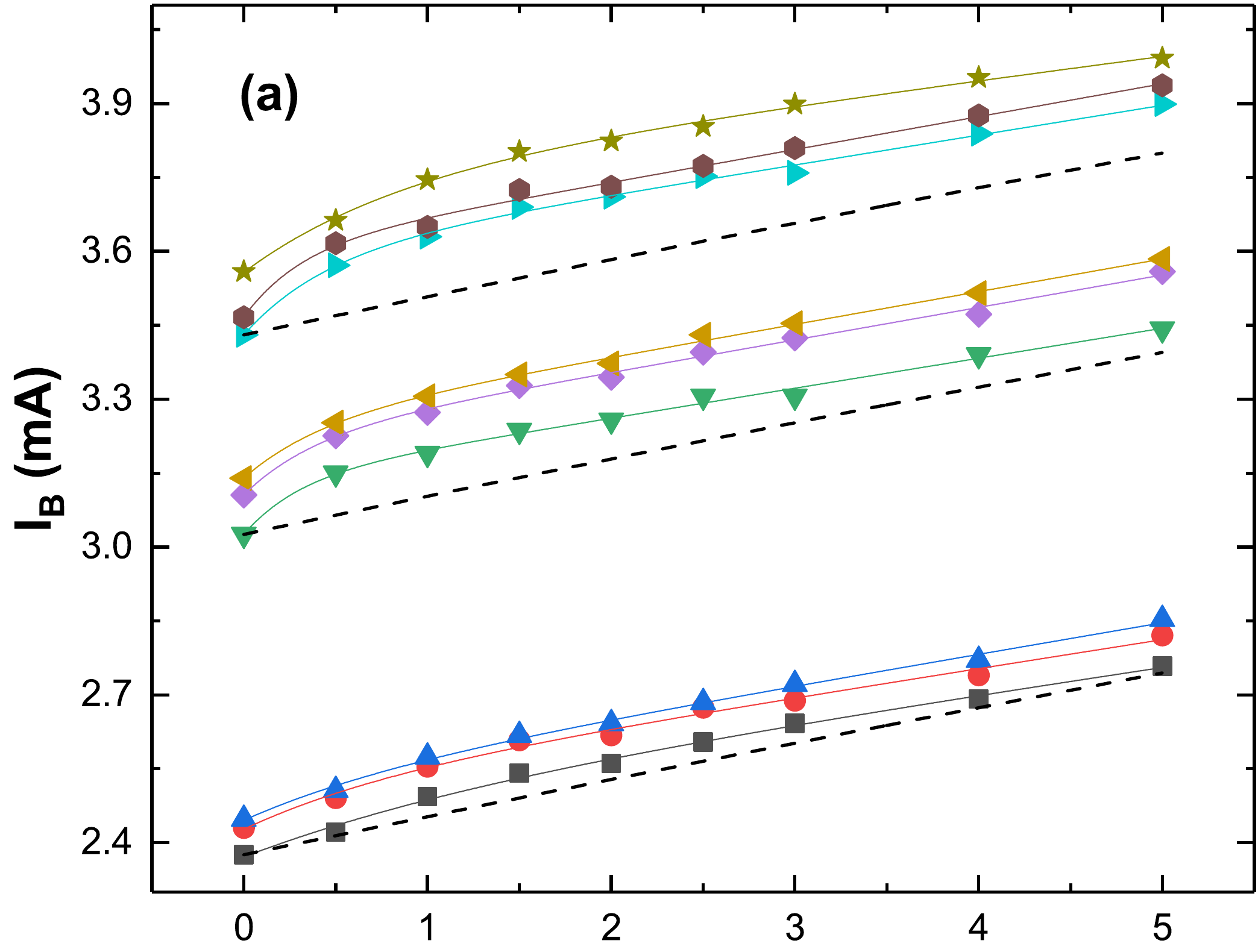}\\
\includegraphics[width=0.85\linewidth]{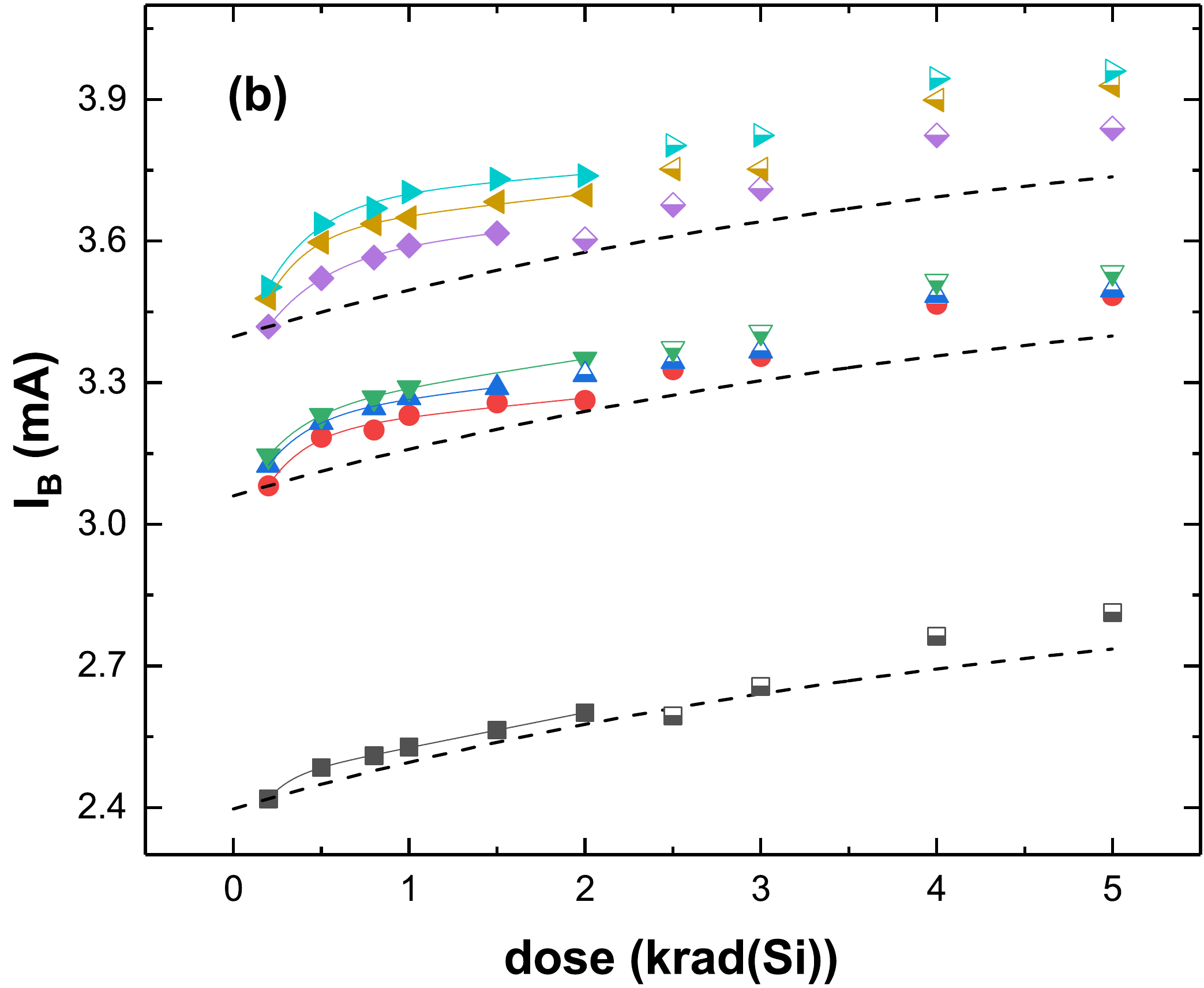}
\caption{The $\gamma$-ray dose and initial base current dependence of the practical damages (dot) and fitting curves (solid curves) of 3DK9D transistors for 10 rad/s (a) and 10 mrad/s (b) dose rates. The summed damages are shown by the dashed lines. 
Reproduced from Song, Y. \emph{et al.}, \emph{ACS Applied Material \& Interfaces} \textbf{2020}, \emph{12}, 29993-29998. Copyright 2020 American Chemical Society.
}\label{fig:SD-HDR}
\end{figure}

{\bf Practical damages and behavior of the negative ISE.}
The $\gamma$-ray response of the $3^{rd}$-$8^{th}$ chips at low dose rate with different $D_0$’s are displayed in Figs.~\ref{fig:TIDafterDDradiation} (b)-(d). It is seen that, the practical damages of all 24 transistors are smaller than their corresponding summed damages. This means that we obtain a clear negative ISE in PNP transistors. Remarkably, the practical damages display `tick’-like curves: the input-bias currents first exponentially {decrease} for relatively small $\gamma$-ray dose and then linearly increase for relatively large $\gamma$-ray dose. Moreover, the slope of the linear increase is obviously smaller than that of the pure ID, $k_0^L$.
The $11^{th}$-$16^{th}$ chips display similar negative ISEs at high dose rate $\gamma$-ray irradiations, see Ref.~\onlinecite{song2019mechanism}.
To further obtain the $D_0$ and $\gamma$-ray dose rate dependences of the ISE, we plot six typical curves in Fig. 5, from which we can see that, the exponential decay of the practical damage at small $\gamma$-ray dose becomes stronger when the dose rate decreases; while the decrease of the slope at large $\gamma$-ray dose becomes bigger when the initial DD increases. These regular behaviors, especially the dose rate dependence of the ISE, cannot be explained by the traditional Coulomb interaction mechanism.

\subsection{Fluence-dependent sublinear dose curves of NPN transistors}

The practical neutron/$\gamma$-ray damage (the base currents) of the 3DK9D NPN transistors are shown in Fig. 6 (a) as a function of the $\gamma$-ray dose at a dose rate of 10 rad/s (see the dots). 
Besides, the summed damages for three typical samples are obtained from a direct sum of their DDs and the average ID of five extra samples (see the dashed curves). 
From the results it is clear that a positive ISE arises for the NPN transistors. From Fig. 6 (a) it is also seen that, the practical damages increase sublinearly as the ionization dose increases; meanwhile, the positive ISE becomes larger when the initial DD increases. 
The practical and summed damages are shown in Fig. 6 (b) for the low dose rate (10mrad/s) case. A second stage arises when the dose exceeds 2 krad. However, the two different stages are found to show similar dependences on the $\gamma$-ray dose and neutron fluence as the high dose rate case.

To sum up, we have designed a variable neutron-fluence and $\gamma$-ray-dose irradiation experiment to explore the behaviors of the negative and positive ISEs in PNP and NPN transistors. The results show that both ISEs display regular dependences on the $\gamma$-ray dose, dose rate, and neutron fluence. 
The obtained features are quite different for PNP and NPN transistors.

\section{Mechanism and analytical model of the positive ISE in NPN transistors}

In all previous works,~\cite{Barnaby2001proton,Barnaby2002Analytical,Li2012Synergistic,
Li2012Simultaneous,Li2015SynergisticEffect}
the ISEs were attributed to the influence of the positively charged OTs accumulated in silica on the charge density in irradiation damaged silicon, see Fig. 1 (a) for NPN transistors and (b) for PNP transistors, respectively. The different polarity of the ISE in PNP and NPN transistors could be explained by this mechanism. However, as indicated above, the dose rate dependence of the effect cannot be explained by the Coulomb interaction mechanism. Besides, the above obtained `tick’-like and sublinear dose dependence of PNP and NPN transistors, respectively, and their strong fluence dependence are also hard to understand from the theory. These behaviors imply that a new underlying mechanism for the ISEs in silicon bipolar transistors should be developed.

We have systematically investigated this basic phenomena from the point of view of \emph{ionization-irradiation-induced evolution of displacement defects in Si}.~\cite{song2020defect,song2019mechanism} The primary consideration is that, the amplitude of the SRH recombination current in the bipolar transistors is in direct proportion to the concentration of the displacement defects in Si; while these defects can evolve under ionization irradiation, due to carrier-enhanced defect diffusion~\cite{kimerling1975role,kimerling1976new,bar1984electronic,Bar-Yam1984,Baraff1984Migration,car1984microscopic}
and recombination-enhanced defect reactions.~\cite{lang1974observation,weeks1975theory,kimerling1978recombination}
The diffusion of displacement defects in Si can be enhanced by alternating capture and lose the ionization-irradiation-induced electrons.~\cite{bar1984electronic,Bar-Yam1984,Baraff1984Migration,car1984microscopic}
On the other hand, the defect reactions can be enhanced by energy liberated upon non-radiative recombination.~\cite{weeks1975theory}
In p-type silicon, $\rm V_2$ is the only neutron-induced active defect, while in n-type silicon, both $\rm V_2$ and VO are neutron-induced active defects, see Fig. 7 (a). In this section, we will first introduce the simpler case, i.e., the understanding and modeling of the ISE in NPN transistors with p-type silicon as the channel.~\cite{song2020defect} Then in the next section we further introduce the case of PNP transistors with n-type silicon as the channel.~\cite{song2019mechanism}

\begin{figure}[!t]
\centering
\includegraphics[width=\linewidth]{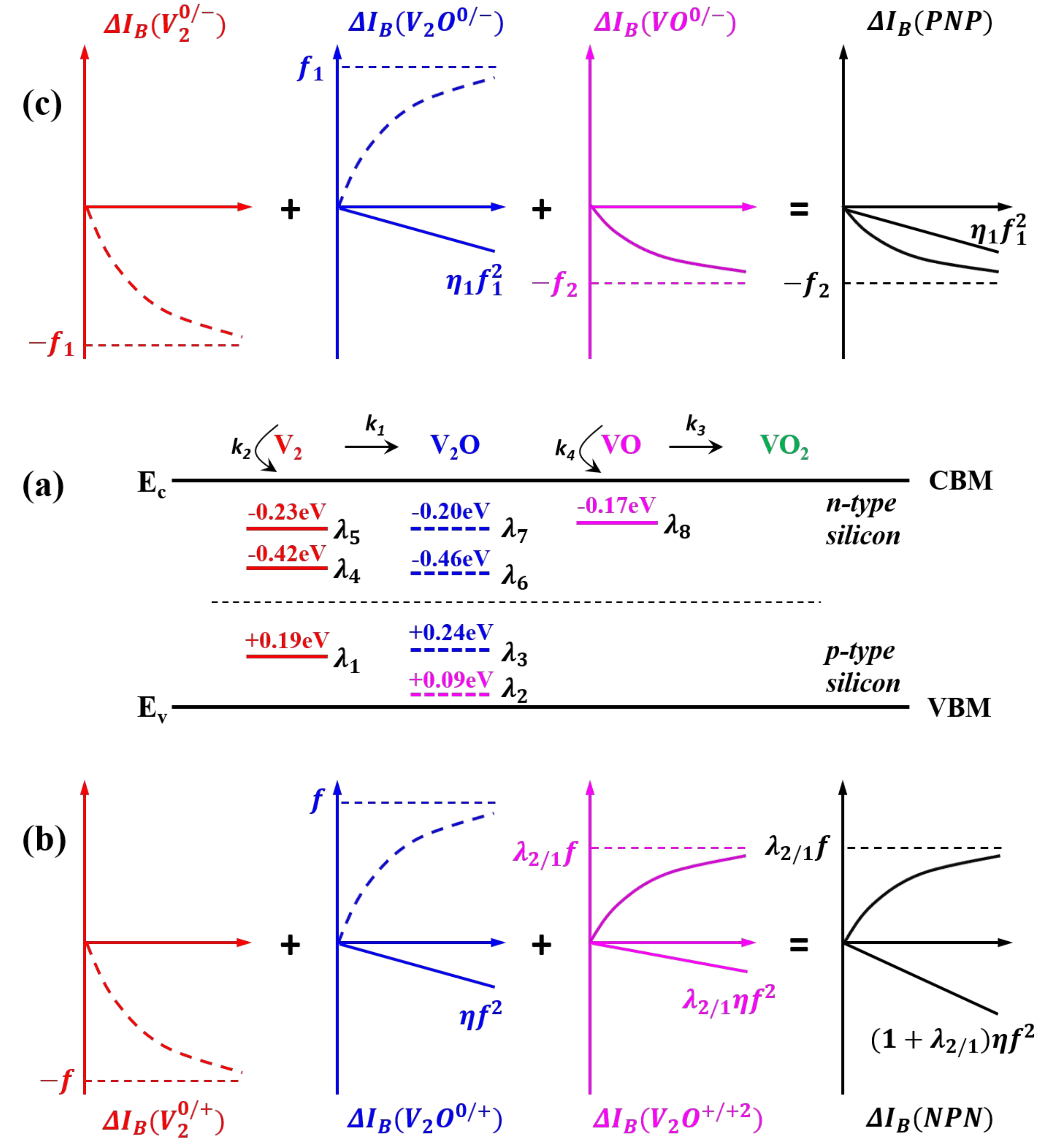}
\caption{Schematic diagram of the proposed defect evolution mechanisms of the ISE. (a) The energy levels of the displacement defects in Si. Bottom-half: in p-type silicon the $\gamma$-ray irradiation can impel the transformation (with a rate of $k_1$) and annihilation (with a rate of $k_2$) of the neutron-irradiation-induced $\rm V_2$ defects. Upper-half: in n-type silicon the evolutions can also happen for VO defects (with a rate of $k_3$). (b) The predicted dose and fluence dependence of the base current in NPN transistors that can be obtained in experiments, together with its compositions due to the evolutions of {$\rm V_2^{0/+}$, $\rm V_2O^{0/+}$, and $\rm V_2O^{+/+2}$} in p-type Si. (c) The predicted dose and fluence dependence of the base current in PNP transistors, together with its compositions due to the evolutions of {$\rm V_2^{0/-}$ ($\rm V_2^{-/-2}$), $\rm V_2O^{0/-}$ ($\rm V_2O^{-/-2}$), and $\rm VO^{0/-}$} in n-type Si.
}\label{fig:mechanism}
\end{figure}

The defect dynamics in p-type silicon under consideration is a $\gamma$-ray induced transformation reaction from $\rm V_2$ to divacancy-oxygen complex ($\rm V_2O$) and a $\gamma$-ray induced annihilation reaction between $\rm V_2$ and self-interstitials ($\rm Si_i$),~\cite{song2020defect} see the bottom half of Fig. 7 (a). Based on these mechanisms, we first derive an analytical model of the defect concentrations as a function of the $\gamma$-ray dose and neutron fluence (Sec. III A). Then we relate the defect concentrations to the base current of the NPN transistors to obtain a model for the ISE (Sec. III B). We at last {test} the proposed model through the variable-dose and -fluence data as shown in Sec. II C (Sec. III C).

\subsection{Defect evolution model in p-type silicon}

As identified by Deep Level Transient Spectroscopy (DLTS),~\cite{lang1974fast,lang1974deep} the main defect type in neutron irradiated p-type silicon is donor-like divacancies ($\rm V_2^{0/+}$),~\cite{cheng19661,lee1976epr,lindstrom1999vibrational} see Fig. 7 (a).
We notice that, the diffusion of $\rm V_2$ will be greatly enhanced when the damaged Si samples are further subjected to a $\gamma$-ray irradiation.~\cite{song2020defect}
The microscopic picture of such a carrier-enhanced defect diffusion is as followings. 
The $\gamma$-ray-induced charge carriers will shift the quasi-Fermi energy in the Si samples. Accordingly, the $\rm V_2$ defects can change from one charge state at its equilibrium site to another that is not at equilibrium.~\cite{bar1984electronic,Bar-Yam1984,Baraff1984Migration,car1984microscopic} 
The latter will easily diffuse to its equilibrium site without having to overcome a diffusion barrier. These behaviors largely enhance the diffusion of $\rm V_2$ defect. 
Besides, many oxygen interstitials ($\rm O_i$), which are electrically-inactive, have been introduced in the Si sample in its growth process.
The moving $\rm V_2$ defects will be trapped by $\rm O_i$ where they react to form another defect $\rm V_2O$ that is {electrically-}active,~\cite{mikelsen2005kinetics,markevich2014donor} 
\begin{subequations}
\begin{equation}
\rm V_2 + O_i \rightarrow V_2O,
\end{equation}
see Fig. 7 (a).

On the other hand, the energy liberated upon nonradiative recombination will be converted into vibrational energy that can be utilized to promote the defect reactions, which include the dissociation of clusters and re-emission of self-interstitials in damaged Si.~\cite{lang1974observation,weeks1975theory,kimerling1978recombination} The charge-enhanced defect diffusion also applies to the emitted self-interstitials.
The moving divacancies and self-interstitials can collide and annihilate each other~\cite{song2020defect}
\begin{equation}
\rm V_2 + 2Si_i \rightarrow 0.
\end{equation}
\end{subequations}

The rate equation of $\rm V_2$ can be derived from Eq. (2). The result is $\rm \partial V_2(t)/\partial t = -k_1 O_i V_2(t) - k_2 Si_i V_2(t)$, where $k_1$ and $k_2$ are the transformation and annihilation reaction velocities in Eqs. (2a) and Eq. (2b), respectively, see Fig. 7 (a).  
To simplify the calculations, we replace $V_2(t)$ and $\rm Si_i$ in the last term by their initial concentrations, which are directly related to the irradiation fluence of neutron, $F=V_2(0)$. 
This replacement can be done because the annihilation rate $k_2$ is relatively small~\cite{mikelsen2005kinetics,markevich2014donor} and the concentration of $\rm Si_i$ is directly proportion to the total amount of vacancies. The above equation becomes
\begin{subequations}
\begin{equation}
\partial V_2(t)/\partial t = -\kappa_1 V_2(t) - \kappa_2 F^2,
\end{equation}
where $\kappa_1=k_1 O_i$ and $\kappa_2=k_2 Si_i/V_2$ are effective transformation and annihilation velocities.
The accompanying rate equation for $\rm V_2O$ defects reads
\begin{equation}
\partial V_2O(t)/\partial t = \kappa_1 V_2(t).
\end{equation}
\end{subequations}
The initial conditions for the two defects are $\rm V_2(0)$ $=F$ and $\rm V_2O(0)$$ = 0$. 
Solving the resulting coupled equations with these conditions, we obtain the $\gamma$-ray dose dependence of the concentration of $\rm V_2$ and $\rm V_2O$ 
\begin{subequations}
\begin{equation}
V_2(D) = (F+\kappa_1^{-1}\kappa_2F^2) (e^{-\kappa_1 R^{-1}D}-1) +F,
\end{equation}
\begin{equation}
V_2O(D) = (F+\kappa_1^{-1}\kappa_2F^2) (1-e^{-\kappa_1 R^{-1}D}) - \kappa_2F^2 R^{-1}D.
\end{equation}
\end{subequations}
Here we have changed $\partial/\partial t$ to $R\partial/\partial D$, where $R$ is the dose rate. From the pair of solutions it is clear that, the concentration of $\rm V_2$ is an exponential decay function of the $\gamma$-ray dose, while the concentration of $\rm V_2O$ is both an asymptotic growth function and a linear decay behavior of the $\gamma$-ray dose.~\cite{song2020defect} The neutron fluence dependence of the defect concentrations is indicated by the factor $F$ in Eq. (4).~\cite{song2020defect}

{\bf Comparison with the annealing experiments.}
Evolution of displacement defects also happens in traditional annealing process induced by elevated temperature. However, the self-interstitial emission mechanism described in Eq. 2 (b) is absent for this case,~\cite{kimerling1978recombination,pellegrino2001annealing} as the energy is not enough to decompose the clusters.
As a result, the defect evolution is only due to defect diffusion and transformation induced by elevated temperatures, as described by Eq. (2a). Then, the defect concentrations as a function of the annealing time ($t$) reduces to
\begin{subequations}
\begin{equation}
V_2(t) = F e^{-\kappa_1 R^{-1}t},
\end{equation}
\begin{equation}
V_2O(t) =  F (1-e^{-\kappa_1 R^{-1}t}).
\end{equation}
\end{subequations}
These results are the same as those in Ref.~\onlinecite{mikelsen2005kinetics}. It is noticed that, here the increase of $\rm V_2O$ equals exactly to the decrease of $\rm V_2$, and we will not observe a linear decay term of the defect concentration as described in Eq. (4b).

\subsection{Recombination current model of NPN transistors}

To show the behavior of the ISE in NPN transistors, we relate the defect concentrations in p-type Si to the SRH base current in the NPN transistor, which are found to be in direct proportion to each other.~\cite{pierret1987advanced, adell2014dose}
The donor levels of the defects in p-type Si have been identified by DLTS.~\cite{mikelsen2005kinetics,markevich2014donor} 
As shown in Fig. 7 (a), there is one active donor level for the $\rm V_2$ defect, $\rm V_2^{0/+}$ at $E_v+0.19$ eV, and two active donor levels for the $\gamma$-ray induced $\rm V_2O$ defect, $\rm V_2O^{0/+}$ at $E_v+0.24$ eV and $\rm V_2O^{+/+2}$ at $E_v+0.09$ eV.
The contribution factors of each defect charge state to the base current can be denoted as $\lambda_1$, $\lambda_2$, and $\lambda_3$, see Fig. 7 (a). Note, the same contribution factor of {$\lambda_1=\lambda_3$} is often assumed~\cite{ganagona2014transformation} as $\rm V_2^{0/+}$ and $\rm V_2O^{0/+}$ have almost the same donor levels and hole capture cross sections.~\cite{srour1973radiation,myers2008model,liu2015radiation} 
The $\gamma$-ray-induced changes of the base current corresponding to the evolution of defects in Eq. (4) read~\cite{song2020defect}
\begin{subequations}
\begin{equation}
\Delta I_B^{V_2^{0/+}}(D,f) = (f+\eta f^2) (e^{-D/D_c} - 1), 
\end{equation}
\begin{equation}
\Delta I_B^{V_2O^{0/+}}(D,f) =  (f+\eta f^2) (1-e^{-D/D_c}) - \eta f^2D/D_c,
\end{equation}
\begin{equation}
\Delta I_B^{V_2O^{+/+2}}(D,f) =  \lambda_{21} \Delta I_B^{V_2O^{0/+}}(D,f).
\end{equation}
\end{subequations}
Here the measurable initial DD due to the neutron-irradiation-induced $V_2$ is denoted as $f=\lambda_1 F$, the ratio between the $\gamma$-ray-irradiation-induced annihilation and transformation velocities is denoted as $\eta=\kappa_1^{-1}\kappa_2/\lambda_1$, $D_c=\kappa_1^{-1} R$ is a characteristic dose of the transformation from $\rm V_2$ to $\rm V_2O$, and $\lambda_{21}=\lambda_2/\lambda_1$ is a ratio between the contribution factors of the first and second donor levels of $\rm V_2O$.
The total change of the base current is the sum of the above three components, which reads~\cite{song2020defect}
\begin{equation}
\Delta I_B (D,f) = \lambda_{21}(f+\eta f^2) (1-e^{-D/D_c})  - (1+\lambda_{21})\eta f^2D/D_c.
\end{equation}
This is the desired analytical model of the positive ISE in NPN transistors, which is based on the ionization-irradiation-induced dynamics of displacement defects in p-type silicon and shows the explicit dependence of the applied $\gamma$-ray dose and initial DD (neutron fluence). These four current profiles are plotted in Fig. 7(b) by using different colors.

The derived model predicts that, the positive ISE of NPN transistors would possess one asymptotic growth term and one linear decay term as a function of the $\gamma$-ray dose,~\cite{song2020defect} see the black curves in Fig. 7 (b).
The asymptotic growth term of the ISE stems from the ionization-irradiation-induced asymptotic growth of the $\rm V_2O^{+/+2}$ defect, see the magenta solid curves in Fig. 7 (b). 
The linear decay term of the ISE results from the two ionization-irradiation-induced linear decaying components of $\rm V_2O^{0/+}$ and $\rm V_2O^{+/+2}$, see the blue and magenta straight solid lines in Fig. 7 (b).
As shown by the red and blue dashed curves in Fig. 7 (b), the remaining asymptotic growth of $\rm V_2O^{0/+}$ and the exponential decay of $\rm V_2^{0/+}$ cancel each other. 
The reason is the one-to-one transformation between $\rm V_2$ and $\rm V_2O$ defects and the same contribution factors of their first donor levels.
The proposed model also predicts an almost linear dependence of the magnitude of the asymptotic growth term on the initial DD ($f$), and a quadratic dependence of the slope of the linear decay term on the initial DD.~\cite{song2020defect} 
The reasons are the one-to-one correspondence between $\rm V_2$ and $\rm V_2O$ in the transformation and the binary behavior of interstitials and vacancies in the annihilation.~\cite{song2020defect}
The proposed defect dynamics model also predicts a dose-rate dependence of the two independent synergistic terms.

\begin{figure}[!t]
\centering
\includegraphics[width=0.86\linewidth]{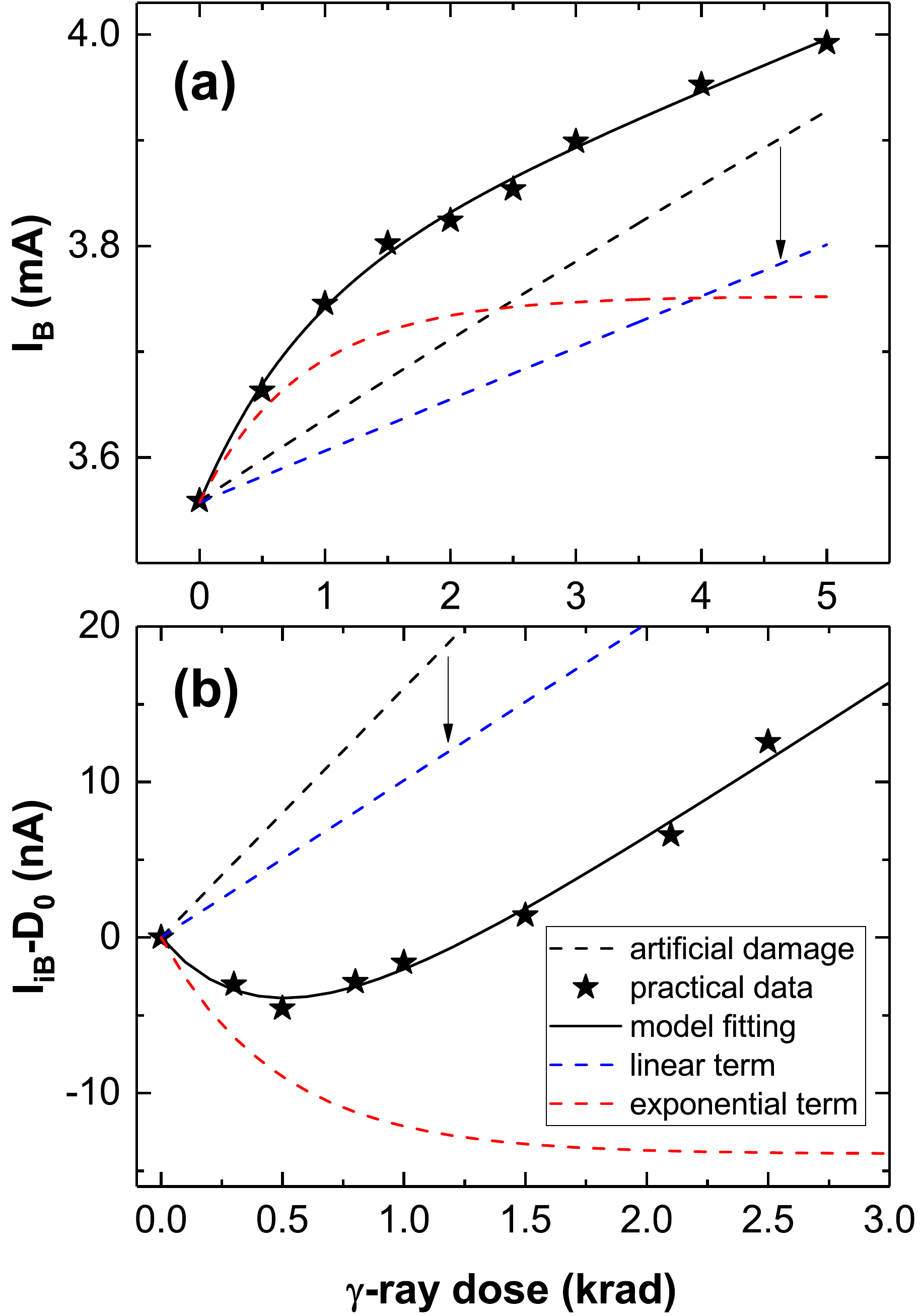}
\caption{The linear (blue dashed) and exponential (red dashed) terms of the practical damage (black dot and solid) for a typical NPN (a) and PNP (b) transistor.
For both cases, the linear term is smaller than the average pure ID (black dashed), indicating the presence of a linear decay term in the ISE.
Reproduced from Song, Y. \emph{et al.}, \emph{ACS Applied Material \& Interfaces} \textbf{2020}, \emph{12}, 29993-29998. Copyright 2020 American Chemical Society.
Reproduced from Song, Y. \emph{et al.}, \emph{ACS Applied Electronic Materials} \textbf{2019}, \emph{1}, 538–547. Copyright 2019 American Chemical Society.
}\label{fig:components}
\end{figure}

\subsection{Experimental verification and parameter extraction of the model of NPN transistors}

To testify the above proposed model for NPN transistors, we try to fit the obtained data in Sec. II.C. by using a fitting model
\begin{equation}
\Delta I_B^j = k_j D + A_j ( 1 - e^{-D/D_c^j} ),
\end{equation}
where $j$ denotes the sample index.
The first term in the fitting model contains both the linear decay term in the proposed physical model and the IDs of the samples. 
As the ID varies among different samples, the physical model, Eq. (7), cannot be directly used. In stead, the presence of the linear decay term can be identified by the statistical difference between the extracted $k_j$ and $k_0$ (the slopes of pure ID). 
The correspondence between the fitting and physical parameters is $k_j-k_0=-(1+\lambda_{21})\eta f^2/D_c$. 
The second term in the fitting model is exactly the asymptotic growth term in the proposed physical model. The correspondence between the parameters is $A_j = \lambda_{21}(f+\eta f^2)$ and $D_c^j = D_c$.
The fitting curves for the high dose rate case are plotted in Fig. 6 (a), which are found to work very well for all 9 samples.
The separated linear component (blue dashed) and asymptotic growth term (red dashed) of a typical sample are shown in Fig. 8 (a). A significant decrease is found in the linear component relative to the average slope of pure ID (black dashed). 
For all 9 samples’ statistical results, see $k_j$ plotted in Fig. 9 (a). 
Similar perfect fitting can be done for the low dose rate case within the first stage, see the fitting curves plotted in Fig. 6 (b). These facts verify the predicted asymptotic growth and linear decay terms in the ISE of NPN transistors, reflecting the proper underlying defect evolution physics in the prediction of dose dependence of the ISE.

\begin{figure*}[!t]
\centering
\flushleft
\begin{minipage}{0.7\textwidth}
\includegraphics[width=0.9\linewidth]{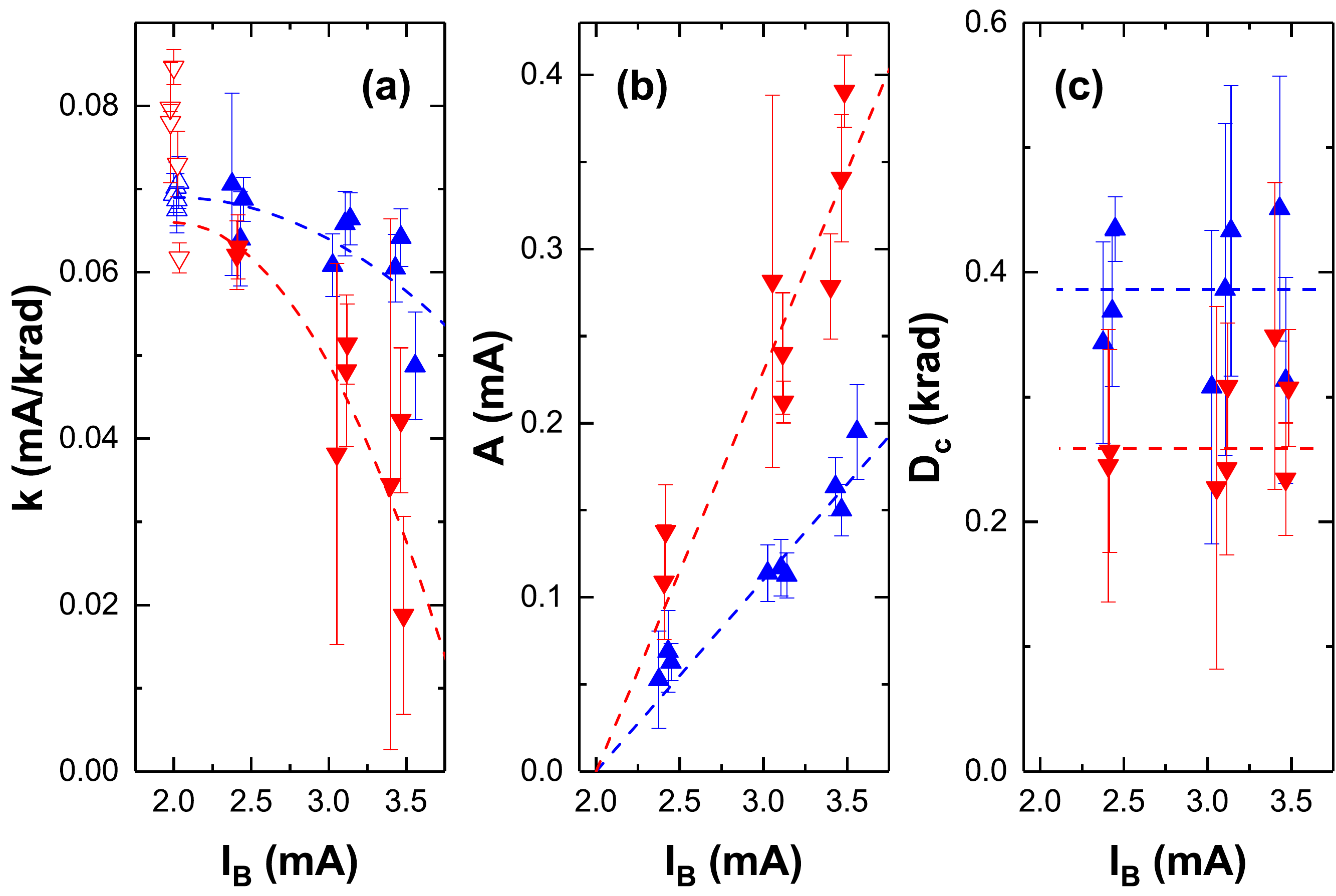}
\end{minipage}%
\begin{minipage}{0.3\textwidth}
\caption{Initial DD dependence of the slope of the linear component (a), as well as the amplitude (b) and characteristic dose (c) of the asymptotic growth term of the ISE for 10 rad/s (blue) and 10 mrad/s (red) $\gamma$-ray irradiation. The dashed curves in (a) and (b) show parabolic and linear guide lines, respectively. 
Reproduced from Song, Y. \emph{et al.}, \emph{ACS Applied Material \& Interfaces} \textbf{2020}, \emph{12}, 29993-29998. Copyright 2020 American Chemical Society.
}\label{fig:SD-HDR}
\end{minipage}
\end{figure*}

To further testify the initial DD dependence of the proposed model, we plot the fitting parameters of Eq. (8) ($k_j$, $A_j$, and $D_c^j$) in Fig. 9 as a function of the base current. 
A decreasing trend is found in Fig. 9 (a) for the slope of the linear term $k_j$ as a function of the initial DD, $D_0$, which starts from the pre-irradiation base current, $\rm I_B^0\approx2mA$. Further investigation shows that a quadratic dependence is satisfied, i.e., $k_j = k_0 - \alpha D_0^2$. This behavior is consistent with the model’s prediction.
Calculations show that $\alpha=5\times10^{-3}\rm mA^{-1} krad^{-1}$ and $17\times10^{-3}\rm mA^{-1} krad^{-1}$ for the high and low dose rate $\gamma$-ray irradiation, respectively.
A linearly increasing trend is found in Fig. 9 (b) for the amplitude of the asymptotic growth term $A_j$ as a function of the initial DD. This feature is also the same as the prediction of the model. The characteristic dose of the asymptotic growth term $D_c^j$ shown in Fig. 9 (c) is insensitive to the DD. These results further confirm the novel initial DD dependence predicted by the proposed model, reflecting the proper defect evolution physics underlying the model.

The significant difference between the red and blue dots in Fig. 9 also reflect the predicted dose-rate dependence of the ISE. 
For a fixed total dose, the carrier-enhanced defect diffusion and recombination-enhanced defect reactions become stronger for decreasing dose rate, which means dramatically increased action time. This fact readily explains the dose rate dependence of the ISE in NPN transistors, which, however, cannot be explained by the Coulomb interaction mechanism.

The physical parameters in Eq. (7) can be calculated from their correspondence to the fitting parameters. The results are shown in Table 1 and some valuable facts can be inferred. The values of $\lambda_{21}$ at different dose rates indicate that the second donor level of $\rm V_2O$ contributes 5 or 10 times less than the first donor level of $\rm V_2O$ and $\rm V_2$. The values of $\eta$ implies that $\eta f \sim 1\times 10^{-3} \ll 1$, which supports $\eta f^2\ll f$ in Eq. (7) hence the linear dependence of the growth term on the initial DD. The dose rate sensitivity of both the dynamic parameters ($D_c=\kappa_1^{-1}R$ and $\eta=\kappa_1^{-1}\kappa_2/\lambda_1$) and the contribution factor ($\lambda_{21}=\lambda_{2}/\lambda_{1}$) of the defects result in the dose rate dependence of the ISE.

\begin{table}[h!]
  \caption{Calculated physical parameters for the ISE in NPN transistors. 
}
  \label{tbl:example}
  \begin{tabular}{llll}
    \hline
    dose rate  & $\lambda_{21}$  & $D_c~\rm (krad)$  & $\eta~\rm (mA^{-1})$\\
    \hline
    10 mrad/s   & 0.23  & 0.26  & $\rm3.9\times10^{-3}$\\
    10 rad/s & 0.11  & 0.38  & $\rm1.5\times10^{-3}$\\
    \hline
  \end{tabular}
\end{table}

\section{Mechanism and analytical model of the negative ISE in PNP transistors}

As discussed in Sec. II B, a negative ISE is observed for PNP transistors. The practical neutron-$\gamma$-ray damage displays an interesting fluence-dependent `tick’-like dose curve. In this section this effect will be understood and modeled in a similar way as that for the NPN transistors. Note that, in n-type silicon, both $\rm V_2$ and VO are neutron-irradiation-induced active defects.

\subsection{ISE due to the evolution of $\rm V_2$ in n-type silicon}

Similar to the p-type silicon case, $\rm V_2$ in n-type silicon as acceptor levels will transform to $\rm V_2O$ due to the above mentioned carrier-enhanced diffusion of $\rm V_2$ under the $\gamma$-ray irradiation,~\cite{kimerling1975role,kimerling1976new,bar1984electronic,Bar-Yam1984,Baraff1984Migration,car1984microscopic} see Fig. 7 (a). 
$\rm V_2$ will also be annihilated by self-interstitials which are emitted due to recombination-enhanced defect reactions.~\cite{lang1974observation,weeks1975theory,kimerling1978recombination}
The ionization dose and neutron fluence dependence of the concentrations of $\rm V_2$ and $\rm V_2O$ are given by Eq. (4).

Different from the p-type silicon case, both $\rm V_2$ and $\rm V_2O$ in n-type silicon have two acceptor levels: $\rm V_2^{-/-2}$ at $\rm E_c-0.23eV$, $\rm V_2^{0/-}$ at $\rm E_c-0.42eV$, $\rm V_2O^{-/-2}$ at $\rm E_c-0.20 eV$, and $\rm V_2O^{0/-}$ at $\rm E_c-0.46eV$,~\cite{mikelsen2005kinetics} see the upper half in Fig. 7 (a).
Denoting the contribution factors of these acceptor charge states to the base current of PNP transistors as $\lambda_4$, $\lambda_5$, $\lambda_6$, and $\lambda_7$, respectively. Note, the same contribution factor of $\lambda_4=\lambda_6$ ($\lambda_5=\lambda_7$) can be assumed, because $\rm V_2^{0/-}$ and $\rm V_2O^{0/-}$ ($\rm V_2^{-/-2}$ and $\rm V_2O^{-/-2}$) have almost the same acceptor levels and electron capture cross sections.~\cite{mikelsen2005kinetics}
Multiplying the defect concentrations, Eq. (4), by these factors, we obtain the corresponding components of the base current in PNP transistors, which are plotted as the red and blue profiles in Fig. 7 (c).
As the exponential decay curve of $\rm V_2^{0/-}$ ($\rm V_2^{-/-2}$) and the asymptotic growth curve of $\rm V_2O^{0/-}$ ($\rm V_2O^{-/-2}$) cancel each other, we can find that the net result due to the ionization-induced evolution of $\rm V_2$ displacement defects in n-type Si is a linear decay term of the ISE in PNP transistors
\begin{equation}
\Delta I_B (D, f_1; V_2\rightarrow V_2O) = - \eta_1 f_1^2D/D_c.
\end{equation}
Here $f_{1}=(\lambda_4+\lambda_5) F$ is the measurable initial DD due to $\rm V_2$ in PNP transistors and $\eta_{1}=\kappa_{1}^{-1}\kappa_{2}/(\lambda_4+\lambda_5)$ is the ratio between the annihilation and transformation reactions for $\rm V_2$ in n-type silicon; these factors are different from the $f$ and $\eta$ factors in Eq. (7) for NPN transistors.

\subsection{ISE due to the evolution of VO in n-type silicon}

Different from p-type silicon, neutron-irradiation-induced VO is also an effective recombination center in n-type silicon, see the upper half in Fig. 7 (a). So, besides $\rm V_2$ the $\gamma$-ray-irradiation-induced evolution of VO should also be considered for PNP transistors. Due to the above mentioned carrier-enhanced diffusion of $\rm VO$~\cite{kimerling1975role,kimerling1976new,bar1984electronic,Bar-Yam1984,Baraff1984Migration,car1984microscopic} 
and recombination-enhanced emission of self-interstitial,~\cite{lang1974observation,weeks1975theory,kimerling1978recombination} the transformation and annihilation reactions of VO will happen
\begin{subequations}
\begin{equation}
\rm VO + O_i \rightarrow VO_2,
\end{equation}
\begin{equation}
\rm VO + Si_i \rightarrow O_i.
\end{equation}
\end{subequations}
According to Ref.~\onlinecite{pellegrino2001annealing},
the transformation reaction happens because the moving VO will be trapped by electrically-inactive $\rm O_i$, which forms an {electrically} non-active defect $\rm VO_2$, see Fig. 7(a). The annihilation reaction happens because the moving vacancy-oxygen complex and self-interstitials can collide and generates electrically-inactive $\rm O_i$. Eqs. (2) and (10) make up the full defect dynamics accounting for the negative ISE in PNP transistors.

Using the same modeling approach as $\rm V_2$ and $\rm V_2O$ in Sec. III B, the concentrations of VO and $\rm VO_2$ can be obtained as functions of the $\gamma$-ray dose and initial DD
\begin{subequations}
\begin{equation}
VO(D) = (G+\kappa_3^{-1}\kappa_4G^2) (e^{-\kappa_3 R^{-1}D}-1) +G,
\end{equation}
\begin{equation}
VO_2(D) = (G+\kappa_3^{-1}\kappa_4G^2) (1-e^{-\kappa_3 R^{-1}D}) - \kappa_4G^2 R^{-1}D.
\end{equation}
\end{subequations}
Here $\kappa_3=k_3 \rm O_i$ and $\kappa_4=k_4 \rm Si_i/VO$ are effective reaction velocities of the transformation and annihilation processes of the defects in n-type Si, respectively, see Fig. 7 (a). The used initial conditions are $\rm VO(0)$ $=G$ and $\rm VO_2(0) = 0$. The results indicate that, similar to the $\rm V_2$ defect, the concentration of VO will exponentially decay under ionization irradiation, and similar to $\rm V_2O$ defect, the concentration of $\rm VO_2$ will asymptotically increase and linearly decay as a function of the $\gamma$-ray dose.

The faded VO defect contributes an acceptor level $\rm E_c-0.17eV$~\cite{svensson1991divacancy} and the generated $\rm VO_2$ defect is non-active in n-type silicon. Denoting the contributing factor of VO to the recombination current as $\lambda_8$, see Fig. 7 (a). The base current in PNP transistors due to VO and $\rm VO_2$ can be solved as a function of the $\gamma$-ray dose and initial DD. The net effect 
is found as an exponential decay of the initial base current ($f_2$) {contributed} by VO
\begin{equation}
\Delta I_B (D,f_2; VO\rightarrow VO_2) = (f_2+\eta_2 f_2^2) (e^{-D/D_c'} - 1),
\end{equation}
see the magenta curve in Fig. 7 (c).
Here $f_{2}=\lambda_8 G$ is the measurable initial DD due to VO defect, $\eta_{2}=\kappa_{3}^{-1}\kappa_{4}/\lambda_8$ is the rate ratio between the annihilation and transformation reactions of VO in Si, and $D_c'=\kappa_{3}^{-1} R$ is a characteristic dose of the transformation reaction of VO in Si.

Summing up Eqs. (9) and (12), we at last obtain the defect dynamics model for the negative ISE of the base current in PNP transistors
\begin{equation}
\Delta I_B (D,f) = - \eta_1 f_1^2D/D_c - (f_2+\eta_2 f_2^2) (1-e^{-D/D_c'}),
\end{equation}
see the black curves in Fig. 7 (c).

Eq. (13) predicts that, there would be two independent terms in the ISEs of the PNP transistors: one linear decay term depending on the $\gamma$-ray dose, which stems from the ionization-irradiation-induced transformation and annihilation of $\rm V_2$ in n-type silicon; and one {exponential decay} term depending on the $\gamma$-ray dose, which stems from the ionization-irradiation-induced transformation and annihilation of VO in n-type silicon.
Eq. (13) also predicts an almost linear dependence of the amplitude of the exponential decay on the initial DD due to VO ($f_2$) and a quadratic dependence of the slope of the linear decay term on the initial DD due to $\rm V_2$ ($f_1$).
Eq. (13) also predicts a dose rate dependence of both terms of the ISE.

\begin{figure}[!b]
\centering
\includegraphics[width=0.86\linewidth]{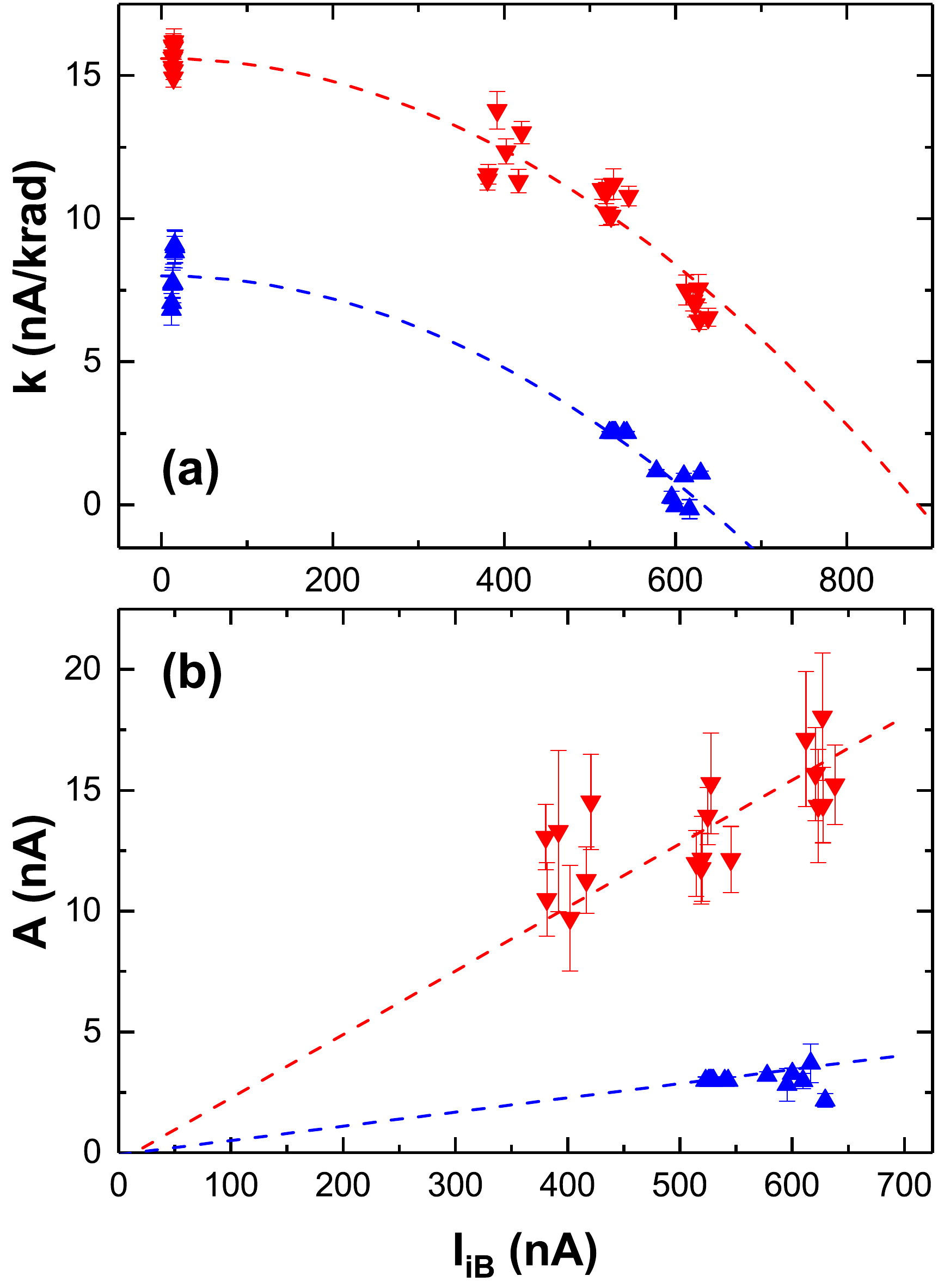}
\caption{
The initial DD dependence of the fitting parameters $k_j$ (a) and $A_j$ (b) in Eq.~(14) for the curves in Fig. 4 at the low (red) and high (blue) dose rate.
}\label{fig:linearterm_annealing}
\end{figure}

\subsection{Experimental verification of the model of PNP transistors}

Following a similar way as the NPN transistor case (Sec. III C), the proposed model, Eq. (13), is testified by the data of PNP devices displayed in Sec. II.B. The fitting model is 
\begin{equation}
\Delta I_B^j =  k_j D - A_j ( 1 - e^{-D/D_c^j} ).
\end{equation}
Again, the first term in the fitting model contains both the linear decay term in the proposed physical model and the unknown linear ID; the presence of the linear decay term can be identified by the statistical difference between $k_j$ and $k_0$. 
The correspondence between the fitting parameters in Eq. (14) and physical parameters in Eq. (13) is $k_j-k_0=-\eta_1 f_1^2/ D_c^j $. The second term in Eq. (14) is exactly the exponential decay term in Eq. (13). The correspondence between the parameters are $A_j = (f_2+\eta_2 f_2^2)\sim f_2$ and $D_c^j = D_c’$. It is noted that $A_j$ is exactly the initial base current due to the VO defects. 
The fitting profiles are plotted in Fig. 5 as solid curves, which are found to work very well for all samples. The dashed curves in Fig. 8 (b) show the linear component and the exponential decay term of a typical fitting curve of the PNP device. 
A significant decrease is found for the linear component (blue dashed) with respect to the average slope of pure ID (black dashed).
These facts verify the predicted linear and exponential decay terms in the ISE of PNP transistors, reflecting the proper defect evolution physics in the prediction of the dose dependence of the ISE in PNP transistors.

The fitting parameters $k_j$ and $A_j$ are plotted in Figure 10 as a function of the base current, to further testify the initial DD dependence of the proposed model. containing the initial DD ($f_1+f_2$). The fitting characteristic dose ($D_c’$) is not sensitive to the initial DD, and is 0.47 krad and 0.50 krad for the low and high dose rates, respectively. Note, these values are larger than those for the $V_2$ defects as shown in tab. 1, reflecting a slower transformation process for VO defects in Si. The data of high dose rate with the neutron fluence of $1\times10^{12}\rm cm^{-2}$ is of low quality and are excluded from the fitting process. It is clear from Fig. 10 (b) that, the amplitude of the {exponential decay} term, $A=f_2$, shows an almost linear dependence on the initial DD, which is started from the pre-irradiation base currents, $\rm I_B^0\approx 14nA$, 
see the dashed guide lines. This dependence is consistent with the prediction of the proposed model. It is also found that for the low (high) dose rate $\gamma$-ray irradiation, the value of $A=f_2$ is only 1/38 (1/170) of the initial DD, $f_1+f_2$. This fact implies that the neutron-induced (annealable) VO defects are much fewer than the neutron-induced $\rm V_2$ defects.
As can be seen in Fig. 10 (a), the slope of the linear term $k_j$ decreases as the initial DD increases. A quadratic dependence is found for the overall trend, $k_j = k_0 - \alpha (f_1+f_2)^2$. Since $f_2$ is much smaller than $f_1$, this result means $k_j = k_0 - \alpha f_1^2$, which is also consistent with the model’s prediction. The value of $\alpha$ is $2\times10^{-5}\rm nA^{-1}krad^{-1}$ for both low and high dose rates.
The above results further verify the predicted initial DD dependence of the ISE in the proposed model, reflecting the proper physics of defect evolution underlying the model.

The significant difference between the red and blue dots in Fig. 10 reflects the dose-rate dependence of the ISE predicted by the model. The reason is that the ionization-induced defect transformation and annihilation becomes stronger for lower dose rate irradiation. This fact readily explains the dose rate dependence of the ISE in PNP transistors, which cannot be explained by the Coulomb interaction mechanism. Different from the NPN transistor case, the physical parameters cannot be calculated from the fitting parameters.

\section{Perspective}

Several important works should be carried out in future investigations on the ISEs of silicon bipolar transistors. The first is the DLTS  characterization of the ionization-irradiation-induced evolution of the displacement defects in n-type and p-type silicon, which can provide direct evidences for the atomistic mechanisms of the ISEs. The second is the explicit dose rate dependence of the ISEs, from which the nature of the two underlying mechanisms, i.e., carrier-enhanced defect diffusion and recombination-enhanced defect reactions, can be further explored. The third and last is the behavior and mechanism of ISE due to \emph{simultaneous} displacement and ionization irradiations. The displacement defect dynamics contains processes of fast generation, short-term annealing, and long-term annealing of displacement defects.~\cite{srour2003review} The above results show that the displacement defects during or after the long-term annealing process can be strongly modified by ionization irradiations.
When the transistors are subjected to simultaneous displacement and ionization irradiations, which is the usual cases, the ionization irradiation can further dramatically alter the short-term annealing of displacement defects.~\cite{gregory1967injection}
The behavior and mechanism of such a \emph{simultaneous ISE} should be an important research subject for both basic researches and practical applications.

\section{Conclusion}

In summary, the irradiation synergistic effect (ISE) is an important physical phenomenon to understand and predict the practical damages of semiconductor devices from the individual displacement and ionization irradiation experiments. In this work, we have made a concise review on the current investigation status of the effect in silicon bipolar transistors. We want to emphasize that, the existing experimental studies of the behavior of the effect are usually limited to fixed displacement fluence and/or fixed ionization dose, while the previous theoretical model encounters physical troubles to explain the dose rate dependence of the ISE.

Here, we describe in detail our recent investigations of the behavior of the effect and introduce our new mechanism of the effect. Our experimental investigation uses a variable-fluence and dose neutron/$\gamma$-ray irradiation setup as well as a large sample size to explore the behavior of the ISE.
Our theoretical investigation adopts a point of view of ionization-irradiation-induced evolution of displacement defects in Si, that is totally different from the Coulomb interaction mechanism in previous works.
We demonstrate that both the positive and negative ISEs in NPN and PNP transistors can be fully and self-consistently explained by
the same basic mechanism, i.e., the ionization-irradiation-induced transformation and annihilation of the displacement-irradiation-induced defects in silicon.
However, as the electrically-active displacement defects and their transformation products are different for p-type and n-type silicon, the ISEs of NPN and PNP transistors show quite different behaviors.
Specifically, the ionization-driven transformation and annihilation of active $\rm V_2$ defect in p-type silicon give arises to a positive ISE in NPN transistors containing an asymptotic growth and a linear decay term;
the same processes of both active $\rm V_2$ and VO defects in n-type silicon give arises to a negative ISE in PNP transistors containing a linear and exponential decay terms.
Both ISEs show a regular dependence on the initial displacement damage of the bipolar transistors, which can also be self-consistently explained by the defect evolution mechanism.
The dose rate problem encountered by the previous Coulomb interaction mechanism can be naturally explained by the present model.

We believe that the uncovered behavior and mechanism of the ISEs will be helpful to predict the practical damage of silicon bipolar transistors. The developed experimental and modeling methods in our works can be extended to the investigation of ISEs in other semiconducting materials and devices.

\section*{Acknowledgments}

This work was supported by the Science Challenge Project under Grant No. TZ2016003-1 and National Natural Science Foundation of China (NSFC) under Grant Nos. 51672023; 11634003; U1930402; 11404300.

\end{document}